\documentclass{svmult}

\usepackage{mathptmx}
\usepackage{helvet}
\usepackage{courier}
\usepackage{makeidx}
\usepackage{multicol}
\usepackage{footmisc}
\usepackage{graphicx}

\begin{document}

\title*{Kermions}

\author{Elizabeth Winstanley}

\institute{Elizabeth Winstanley \at Consortium for Fundamental Physics, School of Mathematics and Statistics,
The University of Sheffield, Hicks Building, Hounsfield Road, Sheffield S3 7RH, United Kingdom. \\
\email{E.Winstanley@sheffield.ac.uk}}

\maketitle

\abstract{
In the framework of quantum field theory in curved space-time,
we study the quantization of a massless fermion field on a non-extremal Kerr black hole.
The key theme in this note is the fundamental difference between scalar and fermion fields for the process of defining quantum states.
In particular, we define two new states for fermions on Kerr which cannot be defined for quantum scalar fields on Kerr.
These two states are the analogues of the standard Boulware and Hartle-Hawking states on a Schwarzschild black hole.
}

\section{Canonical quantization}
\label{sec:intro}

In the canonical quantization of a free field on a curved space-time, an object of fundamental importance is the vacuum state $\left| 0 \right\rangle $.
On a general curved space-time, there is no unique vacuum state.

For a quantum scalar field, the process starts by expanding the classical field $\Phi $ in terms of an orthonormal basis of field modes, which are split into positive frequency modes $\phi _{j}^{+}$ and negative frequency modes $\phi _{j}^{-}$:
\begin{equation}
\Phi = \sum _{j} a_{j}\phi _{j}^{+} + a_{j}^{\dagger } \phi _{j}^{-}.
\label{eq:phisum}
\end{equation}
The choice of positive/negative frequency modes is constrained by the fact that positive frequency modes must have positive Klein-Gordon norm and negative
frequency modes have negative Klein-Gordon norm (by ``norm'', here we mean the inner product of a field mode with itself).
With this restriction, quantization proceeds by promoting the expansion coefficients in (\ref{eq:phisum}) to operators satisfying the usual commutation relations.
The ${\hat {a}}_{j}$ are interpreted as
particle annihilation operators and the ${\hat {a}}_{j}^{\dagger }$ as particle creation operators.
The vacuum state $\left| 0 \right\rangle $ is then defined as that state annihilated by all the particle annihilation operators:
${\hat {a}}_{j} \left| 0 \right\rangle =0$.
The definition of a vacuum state is therefore dependent on how the field modes are split into positive and negative frequency modes, which is restricted for
a quantum scalar field by the fact that positive frequency (particle) modes must have positive norm.

For a fermion field $\Psi $, we again start with an expansion in terms of an orthonormal basis of field modes analogous to (\ref{eq:phisum}):
\begin{equation}
\Psi = \sum _{j} b_{j}\psi _{j}^{+} + c_{j}^{\dagger } \psi _{j}^{-} .
\end{equation}
In this case, both positive and negative frequency fermion modes have positive Dirac norm, so the split of the field modes into positive and negative frequency is much
less constrained for a fermion field compared with a scalar field.
As in the scalar field case, the expansion coefficients are promoted to operators but now they satisfy anti-commutation relations.
The vacuum state $\left| 0 \right\rangle $ is again defined as that state annihilated by all the particle annihilation operators:
${\hat {b}}_{j} \left| 0 \right\rangle =0= {\hat {c}}_{j} \left| 0 \right\rangle $.
Compared with the scalar field case, there is much more freedom in how the vacuum state is defined for a fermion field, because there is much more
freedom in how positive frequency modes can be chosen.

\section{Quantum field theory on Schwarzschild space-time}
\label{sec:schwarz}

Before studying the construction of quantum states on a Kerr black hole, we briefly review the Boulware \cite{Boulware:1975pe} and
Hartle-Hawking \cite{Hartle:1976tp} states on a Schwarzschild black hole.
We emphasize that the construction of these two states is the same for quantum fermion and scalar fields.

\begin{figure}
\subfigures
\begin{minipage}{\textwidth}
\leftfigure[c]{\includegraphics[width=4cm]{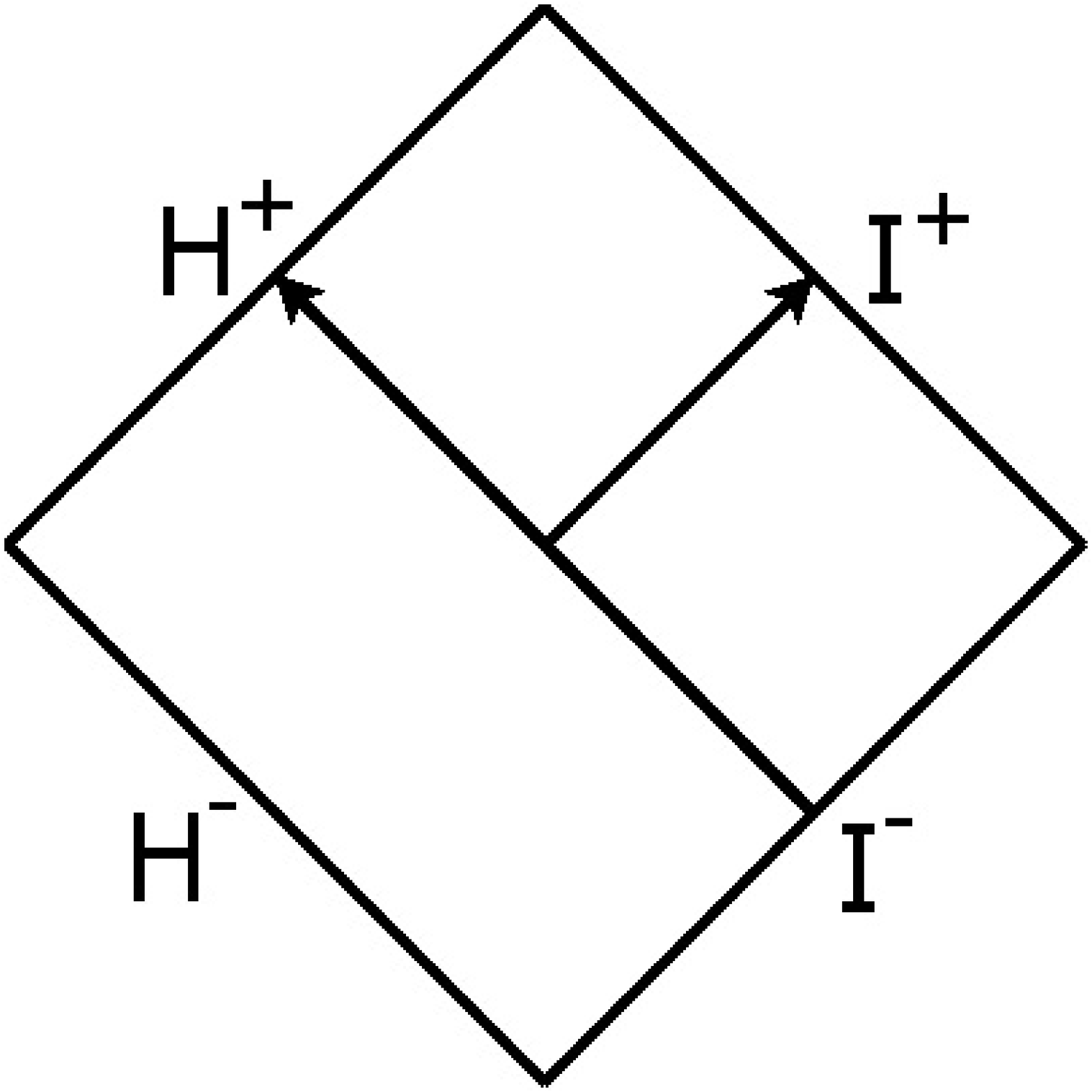}}
\hspace{\fill}
\rightfigure[c]{\includegraphics[width=4cm]{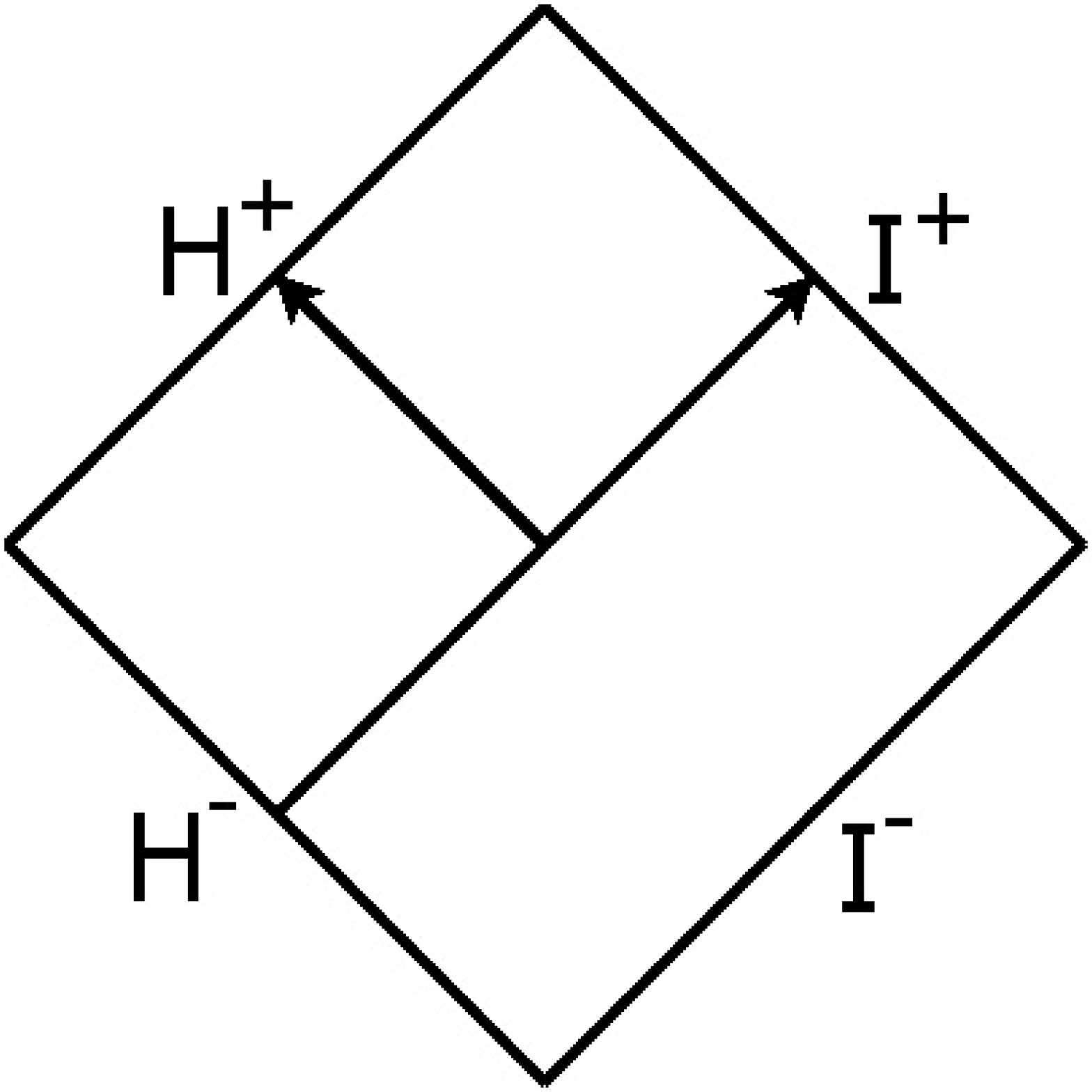}}
\end{minipage}
\leftcaption{``In'' modes: a wave is incident from infinity; part of the wave is reflected back to infinity and part goes down the event horizon of the black hole}
\label{fig:in}
\rightcaption{``Up'' modes: a wave is outgoing from the event horizon; part of the wave is reflected back down the event horizon and part escapes
to infinity}
\label{fig:up}
\end{figure}

Modes for both a massless scalar field and massless fermion field are indexed by the quantum numbers $\omega $, $\ell $ (a total angular momentum quantum number) and $m$ (the azimuthal quantum number).
The quantum number $\omega $ is the frequency of the modes as seen by a static observer either near the event horizon or at infinity.
For scalar field modes, the Klein-Gordon norm is proportional to $\omega $.
We emphasize that {\em {all}} fermion modes have positive Dirac norm.
A suitable basis of field modes consists of the ``in'' and ``up'' modes shown in Figs.~\ref{fig:in}, \ref{fig:up}.

\subsubsection*{Boulware state $\left| B \right\rangle $ \cite{Boulware:1975pe}}

To define this state we choose positive frequency modes as seen by a static observer at infinity.
The resulting vacuum state $\left| B \right\rangle $ contains no particles in the ``in'' and ``up''
modes with $\omega >0$.
This state is as empty as possible at infinity $I^{\pm }$ but diverges on the event horizon $H^{\pm }$.
The unrenormalized expectation value of the stress-energy tensor (UEVSET) for a fermion field in this state is
\begin{equation}
\langle B | {\hat {{\tens {T}}}}_{\mu \nu } | B  \rangle
 =
\frac {1}{2} \sum _{\ell = \frac {1}{2} }^{\infty } \sum _{m=-\ell }^{\ell }
\int _{0}^{\infty } d\omega \left\{
{\tens {T}}_{\mu \nu } \left[\psi _{\omega \ell m}^{{\mathrm {in}}} \right] +
{\tens {T}}_{\mu \nu } \left[ \psi _{\omega \ell m}^{{\mathrm {up}}} \right]
\right\} .
\end{equation}
Here, and in subsequent expressions, we write the UEVSET as a mode sum to show the differences between the
various states considered.
The expression ${\tens {T}}_{\mu \nu } [ \psi _{\omega \ell m}^{{\mathrm {in/up}}} ] $ denotes a classical field mode contribution to the
total UEVSET.

\subsubsection*{Hartle-Hawking state $\left| H \right\rangle $ \cite{Hartle:1976tp}}

In this case we choose positive frequency modes with respect to Kruskal time near the event horizon $H^{\pm }$.
The ``in'' and ``up'' modes with $\omega >0$ become thermally populated with energy $\omega $.
The resulting state $\left| H \right\rangle $ is regular at both the event horizon and infinity; however it is not empty at infinity.
It represents a black hole in thermal equilibrium with a heat bath at the Hawking temperature $T_{H}$.
The UEVSET for a fermion field in this state is
\begin{equation}
\langle H | {\hat {{\tens {T}}}}_{\mu \nu } | H  \rangle
  =
\frac {1}{2} \sum _{\ell = \frac {1}{2}}^{\infty } \sum _{m=-\ell }^{\ell }
\int _{0}^{\infty } d\omega
\tanh \left( \frac {\omega }{2T_{H}} \right)
\left\{
{\tens {T}}_{\mu \nu } \left[\psi _{\omega \ell m}^{{\mathrm {in}}} \right]
+
{\tens {T}}_{\mu \nu } \left[ \psi _{\omega \ell m}^{{\mathrm {up}}} \right]
\right\} .
\end{equation}
For a quantum scalar field, the thermal ``$\tanh $'' factor becomes a ``$\coth $''.

\section{Quantum field theory on Kerr space-time}
\label{sec:kerr}

The Kerr space-time represents a black hole whose event horizon rotates with angular velocity $\Omega _{H}$.
Some key features of the geometry are shown in Figs.~\ref{fig:B}, \ref{fig:H}.
In each figure, the central black region is the interior of the event horizon,
and the axis of rotation runs vertically down the middle of each diagram.
Two further surfaces are important in our later discussion:
\begin{itemize}
\item
The {\em {stationary limit surface}} (the short-dashed surface in Figs.~\ref{fig:B}, \ref{fig:H}) is the surface inside which it is not possible for an observer to remain at rest relative to infinity;
\item
The {\em {speed of light surface}} (the long-dashed surface in Figs.~\ref{fig:B}, \ref{fig:H}) is the surface outside which it is not possible for an observer to co-rotate with the black hole event horizon.
\end{itemize}
A suitable basis of field modes is made up of ``in'' and ``up'' modes as in the Schwarzschild case, indexed by the quantum
numbers $\omega $, $\ell $ and $m$.
A static observer near infinity measures $\omega $ to be the frequency of a particular field mode.
Choosing modes with positive frequency at infinity therefore corresponds to choosing $\omega >0$.
Due to the rotation of the black hole, the corresponding frequency near the horizon is no longer $\omega $ but is shifted by the angular velocity of the black hole to be ${\tilde {\omega }} = \omega - m\Omega _{H}$.
Choosing modes with positive frequency with respect to Kruskal time near the horizon therefore corresponds to thermally populating the ``in'' and ``up'' modes with energy ${\tilde {\omega }}$ rather than $\omega $.
A further complication is that, for scalar fields, the ``in'' modes have positive norm only if $\omega >0$ whereas the ``up'' modes have positive
norm only if ${\tilde {\omega }}>0$.
As previously, for a fermion field all modes have positive norm.

We now describe three quantum states on Kerr space-time, including analogues of the Boulware and Hartle-Hawking states defined above for Schwarzschild black holes.

\subsubsection*{``Past'' Boulware state $\left| B^{-} \right\rangle $ \cite{Unruh:1974bw}}

Based on the norms of the scalar field modes, the natural frequency for the ``in'' modes is $\omega $, while for the ``up'' modes it is
${\tilde {\omega }}$. Choosing $\omega >0$ for the ``in'' modes and ${\tilde {\omega }}>0$ for the ``up'' modes as the definition of
positive frequency leads to the ``past'' Boulware \cite{Ottewill:2000qh,Unruh:1974bw} state, for which the UEVSET for a fermion field is \cite{Casals:2012es}
\begin{equation}
\langle B^{-} | {\hat {{\tens {T}}}}_{\mu \nu } | B^{-} \rangle
 =
\frac {1}{2}\sum _{\ell = \frac {1}{2}}^{\infty } \sum _{m=-\ell }^{\ell }
\left\{
\int _{0}^{\infty } d\omega \,
{\tens {T}}_{\mu \nu } \left[\psi _{\omega \ell m}^{{\mathrm {in}}} \right]
+
\int _{0}^{\infty } d{\tilde {\omega }} \,
{\tens {T}}_{\mu \nu } \left[ \psi _{\omega \ell m}^{{\mathrm {up}}} \right]
\right\} .
\label{eq:pastB}
\end{equation}
Like the Boulware state on a Schwarzschild black hole, the ``past'' Boulware state is divergent on the event horizon of a Kerr black hole, but it is
regular everywhere outside the event horizon.
Unlike the Boulware state on a Schwarzschild black hole, it is not empty at infinity; there is an outgoing flux of radiation at $I^{+}$
\cite{Unruh:1974bw}.
The construction of the ``past'' Boulware state is identical for both scalar and fermion fields.

\subsubsection*{``Boulware'' state $\left| B \right\rangle $  \cite{Casals:2012es}}

\begin{figure}
\subfigures
\begin{minipage}{\textwidth}
\leftfigure[c]{\includegraphics[width=6cm]{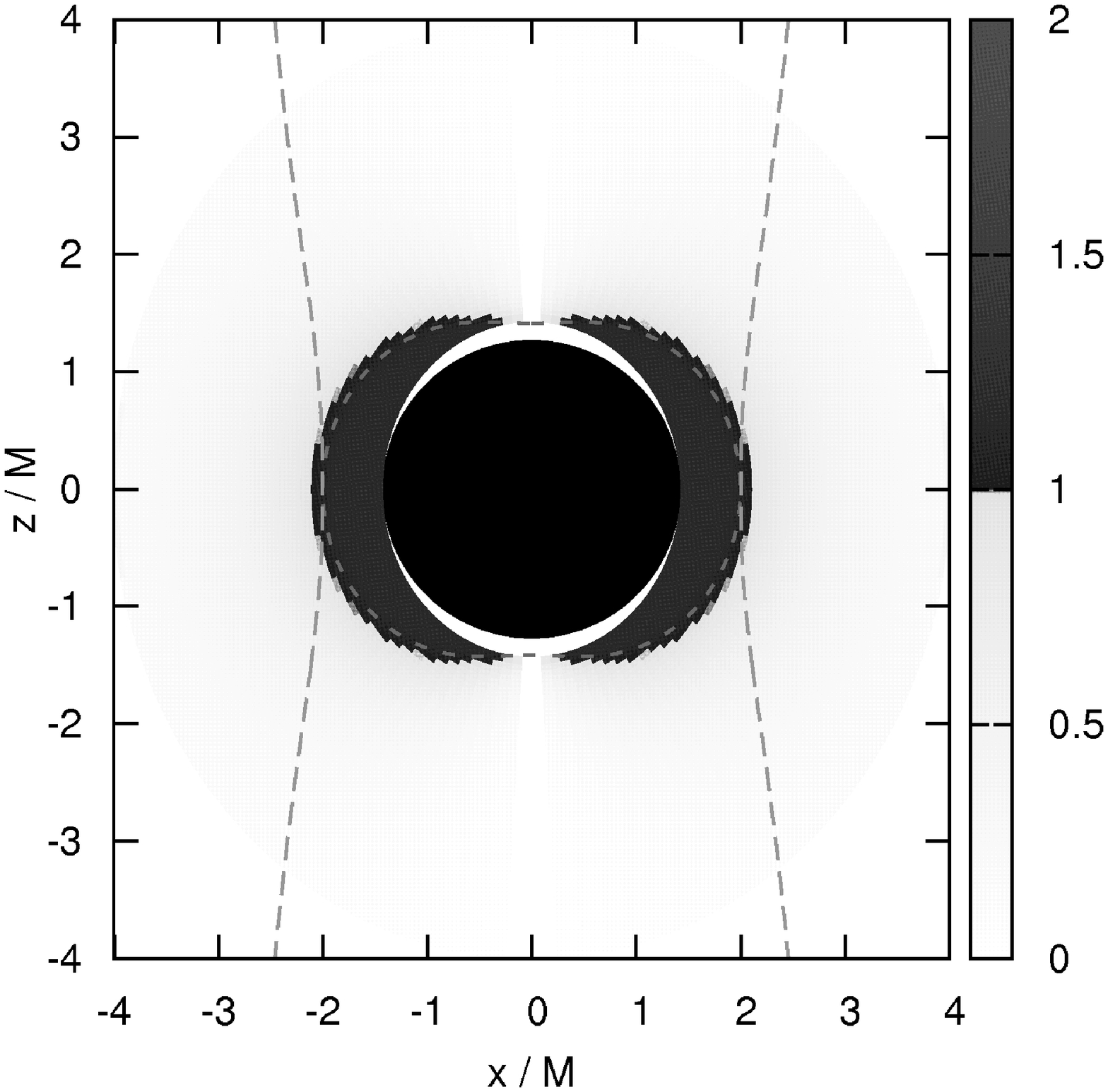}}
\hspace{\fill}
\rightfigure[c]{\includegraphics[width=6cm]{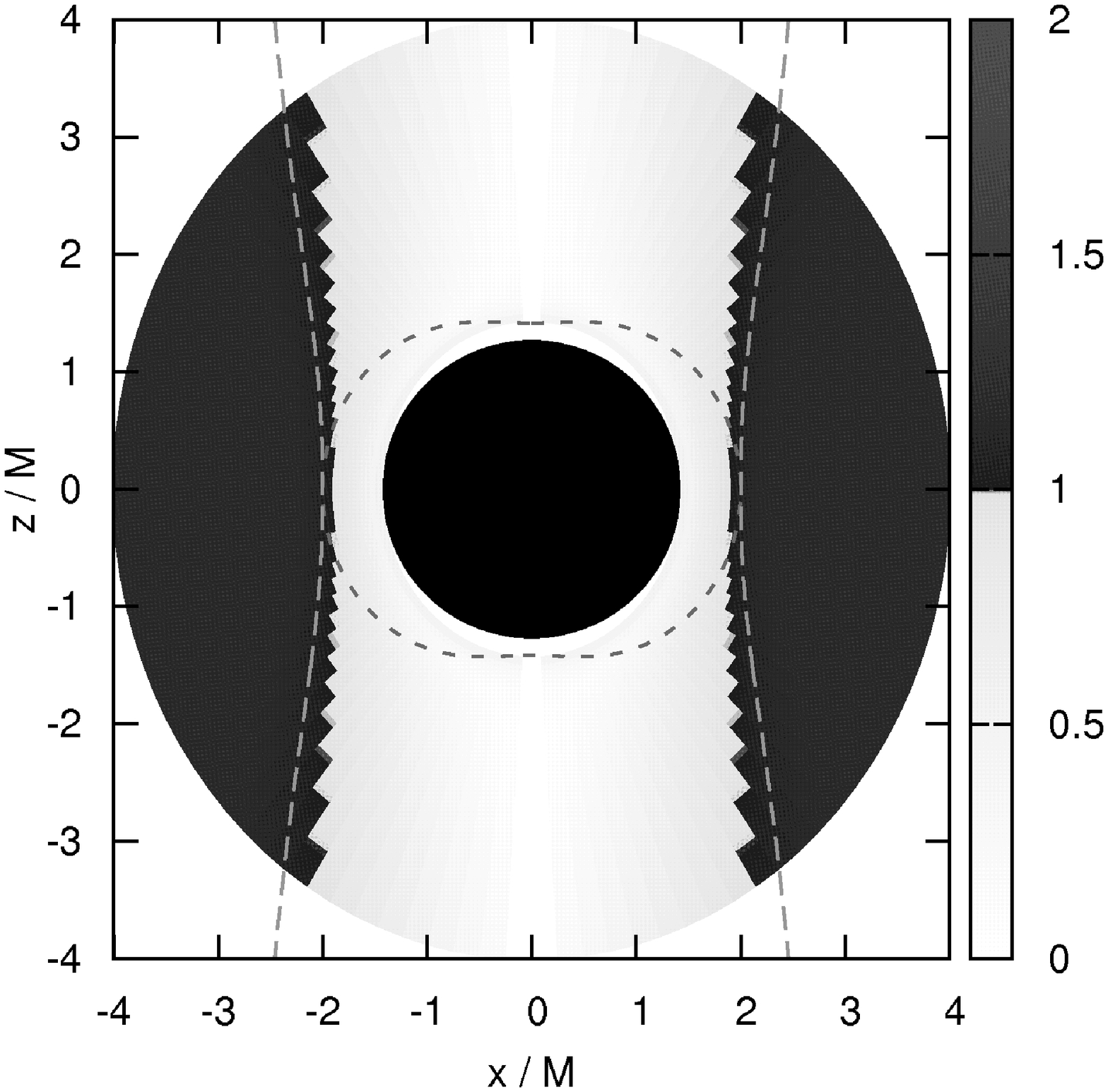}}
\end{minipage}
\leftcaption{Regularity properties of  $\left| B\right\rangle $.
The UEVSET (\ref{eq:pastB}) is subtracted from (\ref{eq:B}). Dark regions indicate where this difference is divergent and light regions where
it is finite.
Taken from \cite{Casals:2012es} }
\label{fig:B}
\rightcaption{Regularity properties of $\left| H\right\rangle $.
The UEVSET (\ref{eq:pastB}) is subtracted from (\ref{eq:H}).
Dark regions indicate where this difference is divergent and light regions where
it is finite.
Taken from \cite{Casals:2012es} }
\label{fig:H}
\end{figure}

We next seek to construct an analogue of the Boulware state on Schwarzschild space-time, which is as empty as possible at both future and past null infinity, $I^{\pm }$.
We would therefore like to choose $\omega >0$ as positive frequency for both ``in'' and ``up'' modes.
For scalar fields there is an immediate problem:  the ``up'' modes have positive norm for ${\tilde {\omega }}>0$, not $\omega >0$.
As a result for scalar fields no state empty at both future and past null infinity $I^{\pm }$ can be defined \cite{Ottewill:2000qh}.
However, for fermion fields all modes have positive norm and in this case we can construct a state, the ``Boulware'' state $\left| B \right\rangle $, by taking $\omega >0$ as the definition of positive frequency for all modes.
The UEVSET for a fermion field in this state is \cite{Casals:2012es}
\begin{equation}
\langle B | {\hat {{\tens {T}}}}_{\mu \nu } | B  \rangle
 =
\frac {1}{2} \sum _{\ell = \frac {1}{2}}^{\infty } \sum _{m=-\ell }^{\ell }
\int _{0}^{\infty } d\omega \left\{
{\tens {T}}_{\mu \nu } \left[\psi _{\omega \ell m}^{{\mathrm {in}}} \right] +
{\tens {T}}_{\mu \nu } \left[ \psi _{\omega \ell m}^{{\mathrm {up}}} \right]
\right\}  .
\label{eq:B}
\end{equation}
To determine where the state $\left| B \right\rangle $ is regular, we subtract from (\ref{eq:B}) the UEVSET (\ref{eq:pastB}),  since the state
$\left| B^{-} \right\rangle $ is regular everywhere outside the event horizon.
It can be seen from Fig.~\ref{fig:B} that this difference in expectation values is regular outside the stationary limit surface of the Kerr black hole, but
diverges in the ergosphere.
We therefore deduce that the state $\left| B \right\rangle $ is also regular outside the stationary limit surface but divergent in the ergosphere.

\subsubsection*{``Hartle-Hawking'' state $\left| H \right\rangle $  \cite{Casals:2012es} }

To define an analogue of the Hartle-Hawking state on Schwarzschild space-time, we would like to construct a state representing a Kerr
black hole in thermal equilibrium with a heat bath at the Hawking temperature $T_{H}$.
Since the frequency of the modes near the horizon is ${\tilde {\omega }}$, this would correspond to thermally populating the ``in'' and ``up''
modes with energy ${\tilde {\omega }}$.
For a quantum scalar field, this cannot be done because the ``in'' modes are defined in terms of frequency $\omega $, not ${\tilde {\omega }}$.

On the other hand, for fermions on Kerr we are able to define a ``Hartle-Hawking'' state for which both the ``in'' and ``up'' modes are thermalized
with energy ${\tilde {\omega }}$.
In this case the UEVSET for the fermion field is \cite{Casals:2012es}
\begin{equation}
\langle H | {\hat {{\tens {T}}}}_{\mu \nu } | H  \rangle
  =
\frac {1}{2} \sum _{\ell = \frac {1}{2}}^{\infty } \sum _{m=-\ell }^{\ell }
\int _{0}^{\infty } d{\tilde {\omega }}
\tanh \left( \frac {{\tilde {\omega }}}{2T_{H}} \right)
\left\{
{\tens {T}}_{\mu \nu } \left[\psi _{\omega \ell m}^{{\mathrm {in}}} \right]
+
{\tens {T}}_{\mu \nu } \left[ \psi _{\omega \ell m}^{{\mathrm {up}}} \right]
\right\} .
\label{eq:H}
\end{equation}
To determine where the state $\left| H \right\rangle $ is regular, we subtract from (\ref{eq:H}) the UEVSET (\ref{eq:pastB}).
It can be seen from Fig.~\ref{fig:H} that this difference in expectation values is regular between the event horizon and the speed-of-light surface, but
diverges outside the speed-of-light surface.
We therefore deduce that the state $\left| H \right\rangle $ is also regular between the event horizon and the speed-of-light surface but diverges outside the speed-of-light surface.

\begin{acknowledgement}
This note discusses work completed in collaboration with Marc Casals, Sam Dolan, Brien Nolan and Adrian Ottewill.
This work was supported by the Lancaster-Manchester-Sheffield Consortium for Fundamental Physics under STFC Grant No.~ST/J000418/1,
by an International Visitor Programme Grant from the Office of the Vice President for Research in Dublin City University and by EU COST Action
MP0905 ``Black Holes in a Violent Universe''.
We thank Victor Ambrus for helpful discussions.
\end{acknowledgement}


\begin{thebibliography}{99}

\bibitem{Boulware:1975pe}
  Boulware, D.~G.:
  Spin 1/2 quantum field theory in Schwarzschild space.
  Phys.\ Rev.\ D {\bf 12}, 350--367 (1975)

\bibitem{Casals:2012es}
  Casals, M.; Dolan, S.~R.; Nolan, B.~C.; Ottewill, A.~C. and Winstanley, E.:
  Quantization of fermions on Kerr space-time.
  Phys.\ Rev.\ D {\bf 87}, 064027 (2013)

\bibitem{Hartle:1976tp}
  Hartle, J.~B. and Hawking, S.~W.:
  Path integral derivation of black hole radiance.
  Phys.\ Rev.\ D {\bf 13}, 2188--2203 (1976)

\bibitem{Ottewill:2000qh}
  Ottewill, A.~C. and Winstanley, E.:
  The renormalized stress tensor in Kerr space-time: general results.
  Phys.\ Rev.\ D {\bf 62}, 084018 (2000)

\bibitem{Unruh:1974bw}
  Unruh, W.~G.:
  Second quantization in the Kerr metric.
  Phys.\ Rev.\ D {\bf 10}, 3194--3205 (1974)

\end{thebibliography}
\end{document}